\documentclass[preprint2,11pt]{aastex}
\usepackage{epsfig}
\usepackage{natbib}
\usepackage{amsmath}
\usepackage{multicol}
\usepackage{color}

\newcommand{\kms}{km~s$^{-1}$}

\DeclareMathAlphabet{\mathpzc}{OT1}{pzc}{m}{it}

\begin{document}
\title{Cosmicflows-3: Cold Spot Repeller?}

\author{H\'el\`ene M. Courtois}
\affil{University of Lyon, UCB Lyon 1, CNRS/IN2P3, IPN Lyon, France}
\and
\author{R. Brent Tully,}
\affil{Institute for Astronomy, University of Hawaii, 2680 Woodlawn Drive, Honolulu, HI 96822, USA}
\and
\author{Yehuda Hoffman}
\affil{Racah Institute of Physics, Hebrew University, Jerusalem, 91904 Israel}
\and
\author{Daniel Pomar\`ede}
\affil{Institut de Recherche sur les Lois Fondamentales de l'Univers, CEA, Universit\'e Paris-Saclay, 91191 Gif-sur-Yvette, France}
\and
\author{Romain Graziani}
\affil{University of Lyon, UCB Lyon 1, CNRS/IN2P3, IPN Lyon, France}
\and
\author{Alexandra Dupuy}
\affil{University of Lyon, UCB Lyon 1, CNRS/IN2P3, IPN Lyon, France}

\begin{abstract}
The three-dimensional gravitational velocity field within $z\sim0.1$ has been modeled with the Wiener filter methodology applied to the {\it Cosmicflows-3} compilation of galaxy distances.  The dominant features are a basin of attraction and two basins of repulsion.  The major basin of attraction is an extension of the Shapley concentration of galaxies.  One basin of repulsion, the Dipole Repeller, is located near the anti-apex of the cosmic microwave background dipole.  The other basin of repulsion is in the proximate direction toward the 'Cold Spot' irregularity in the cosmic microwave background.  It has been speculated that a vast void might contribute to the amplitude of the Cold Spot from the integrated Sachs-Wolfe effect.

\smallskip\noindent
Key words: large scale structure of universe --- galaxies: distances and redshifts
\bigskip
\end{abstract}

\smallskip
\section{Introduction}

All-sky maps of the structure in the universe are increasingly extensive.  Here we use the {\it Cosmicflows-3} (CF3) collection of 18,000 galaxy distances \citep{2016AJ....152...50T} to study velocities that depart from Hubble expansion on a scale of $0.1c$.  The data assembly builds on the 8,000 distances of {\it Cosmicflows-2} \citep{2013AJ....146...86T} primarily with two new sources.  All-sky luminosity-linewidth measurements with infrared Spitzer satellite photometry provide distances to relatively nearby galaxies, giving improved coverage at low galactic latitudes \citep{2012AJ....144..133S, 2014MNRAS.444..527S}.  Of greatest importance for the present discussion, though, is the contribution from the 6dFRSv program \citep{2012MNRAS.427..245M,2014MNRAS.443.1231C,2014MNRAS.445.2677S} of Fundamental Plane distances to galaxies in the south celestial hemisphere, a zone underrepresented in {\it Cosmicflows-2}.

\section{Methods}

Our derivation of three-dimensional peculiar velocity and mass density fields from observed distances follows the Wiener filter methodology \citep{1999ApJ...520..413Z,2009LNP...665..565H,2012ApJ...744...43C} assuming a power spectrum consistent with the Lambda Cold Dark Matter model with the cosmological parameters given by \citet{2009ApJS..180..330K}. This Bayesian prior is highly constrained by data nearby where coverage is dense and accurate but decays to the exigencies of the power spectrum at large distances where data are sparse and have large errors.  The amplitudes of the estimated density and velocity fluctuations are suppressed at large distances due to uncertainties.  However, power in dipole and quadruple terms signal the influence of important structures at the farthest extremities of the observational data.

A familiar way to represent the velocity field is with vectors, either located at specified galaxies or seeded on a grid \citep{1999ApJ...520..413Z}.  Our preference is to use streamlines \citep{2017NatAs...1E..36H}.  A streamline  $\vec{l}(s)$ is computed by integrating  $d \vec{r}(s) = \vec{v}(\vec{r}(s)) d s$.  A streamline can be seeded on a regular grid, or at specified positions such as regions of maxima and minima of the potential.  Integrating the streamlines over many integration steps, flows from low density regions either leave the computational box or converge onto a basin of attraction.  Anti-flow streamlines, the negative of the flow field, proceed from high densities either out of the box or to basins of repulsion \citep{2017NatAs...1E..36H}.

The structures to be discussed are at large distances where the peculiar velocity field can be badly compromised by Malmquist bias \citep{1995PhR...261..271S}.  Our methodology will be discussed at length by Graziani et al. (in preparation).  Briefly, distance errors create an artifact of flows toward the peak in the sample distribution because there are more targets scattering away from the peak than into the peak.  The peak is set by the convolution of increasing candidates with volume and the loss of candidates at large distances.  Except very nearby, the true positions of galaxies is approximately set by their redshifts.  A probability distribution can be calculated for the peculiar velocity of each target from a Bayesian analysis based on the assumption of Gaussian errors constrained by the observed distance and error estimate.  There is the resolvable complexity that errors that are approximately Gaussian in the distance modulus give log normally distributed errors in peculiar velocities \citep{2015MNRAS.450.1868W}.

We evaluate the robustness of our results through comparisons between many constrained realizations \citep{1991ApJ...380L...5H}.  The Wiener filter analysis establishes the minimum variance mean fields of velocity and mass density and constrained realizations then sample the scatter around the Wiener filter mean field \citep{1999ApJ...520..413Z}.
 
\section{Results}

Our Wiener filter analysis, based strictly on peculiar velocities and in ignorance of the distribution of galaxies, produces an intriguing complementary view to structure identified by redshift surveys.  Figure 1 and the companion interactive figure\footnote{The Sketchfab interactive figure can be launched from https://skfb.ly/6sXCY. Selecting on numbers will take the viewer on a pre-determined path.  The model can be manipulated freely with mouse controls.} provide a visual summary based on current information.  The extreme peaks and valleys of the gravitational potential inferred in linear theory from the filtered velocity field are represented by iso-potential surfaces, attractors and repellers distinguished by colors.     
In the top panel, peculiar velocity flow lines start at seeds within the basins of repulsion and end at the major attractors.  In the bottom panel, the flow is inverted and anti-flow streamlines seeded in the basins of attractions converge onto the repellers.

\subsection{Repellers}  

Two repeller sinks are identified in Figure 1.  One was already  apparent in the Wiener filter study of the precursor {\it Cosmicflows-2} dataset and was named the Dipole Repeller \citep{2017NatAs...1E..36H}.  Its location, as computed with  the {\it Cosmicflows-3} dataset, is at galactic glon=$94^\circ$, glat=$-16^\circ$, distance=14,000~\kms, displaced $2^{\circ}$ and 2,000~\kms\ (12\%) closer than found previously.  The cosine of the anti-alignment with the direction of the cosmic microwave dipole \citep{1996ApJ...473..576F} is $\mu=-0.99$; the agreement in direction is $9^{\circ}$.  \citet{2017NatAs...1E..36H} evaluated the uncertainties in the direction and depth of the Dipole Repeller based on constrained realizations and sample cuts with the {\it Cosmicflows-2} dataset and found $\mu=-0.96\pm0.04$ and distance 16,000~\kms.  The new results are in even better agreement with the dipole anti-pole.
It was inferred by Hoffman et al. that the Dipole Repeller accounts for roughly half of the Milky Way motion reflected in the dipole.

\begin{figure*}[h!]
\begin{center}
\includegraphics[scale=.85]{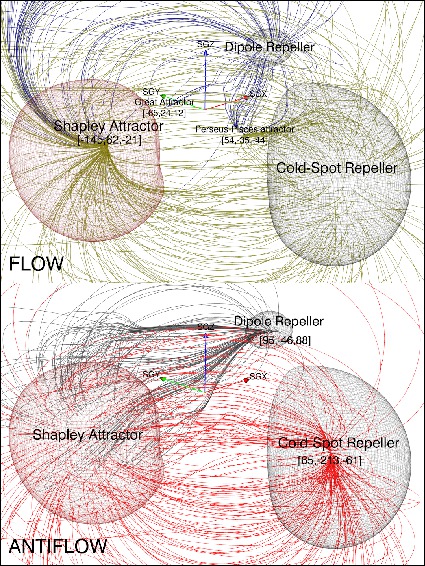}
\caption{Dominant attractors and repellers in the nearby universe.  In the top panel flow streamlines are seeded in the Dipole and Cold Spot repellers, flows respectively blue and green, and converge predominantly onto the Shapley and Perseus-Pisces attractors. In the bottom panel anti-flow streamlines are seeded in the Shapley attractor, with flows in red and black that travel to the Cold Spot and Dipole repellers respectively.  Our home is located by the red, green, blue arrows of length 10,000~\kms\ directed along the supergalactic SGX, SGY, SGZ axes. The locations of the attractors and repellers lie at the local extrema of the potential.  The surfaces that represent the attractors and voids are at symmetric values of the potential field of $\pm 580$. The positions are in supergalactic coordinates and are expressed in units of 100~\kms.}
\label{fig1}
\end{center}
\end{figure*}

\begin{figure*}[h!]
\begin{center}
\includegraphics[scale=.27]{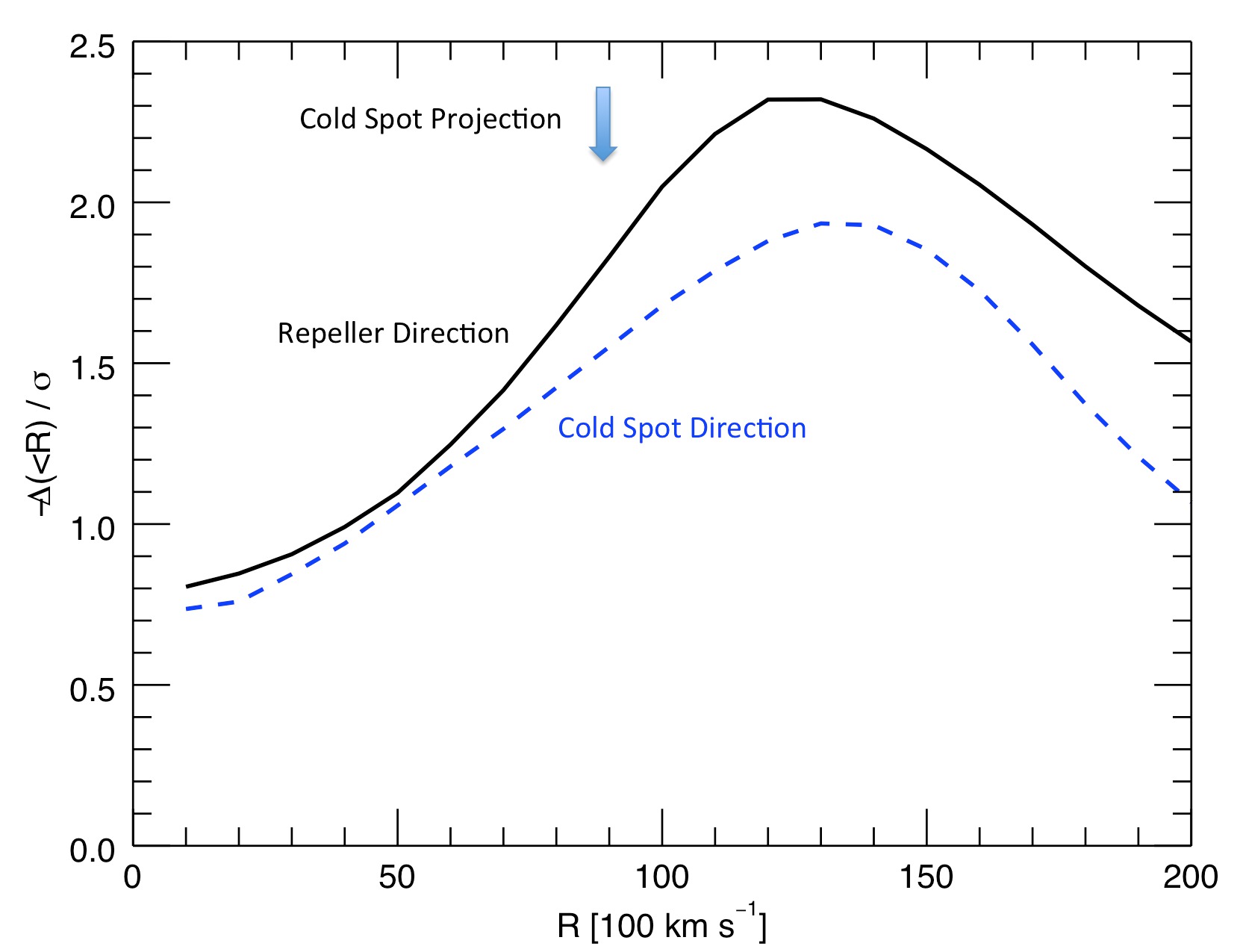}
\caption{Ratio of the Wiener filter estimated underdensity to r.m.s. scatter over an ensemble of 54 constrained realizations in spheres of increasing radius centered on the potential extremum (solid curve) and on a point at the same distance in the direction toward the Cold Spot (dashed curve).  In each case, only the volume within 25,000~\kms\ of the Milky Way is considered.  The distance from the bottom of the basin of the repeller to the intersection with the direction to the Cold Spot is indicated by the arrow.}
\label{fig2}
\end{center}
\end{figure*}

It is the second repeller sink that draws our current attention.  The rough location is glon=168$^\circ$, glat=$-71^\circ$, velocity=23,000~\kms. 
It is in the general region of the negative velocity anomaly in Pisces-Cetus noted by \citet{2014MNRAS.445.2677S}.  We particularly note that it is in the direction of a feature projected against the so-called CMB Cold Spot \citep{2015MNRAS.450..288S}.  The Cold Spot is a fluctuation in the cosmic microwave background at glon$=209^\circ$, glat$= -57^\circ$ with an amplitude that has been argued is difficult to reconcile with the standard $\Lambda$ Cold Dark Matter model \citep{2004ApJ...609...22V,2008MNRAS.390..913C}.  There were early hints of an underdensity of galaxies in the same direction \citep{2007ApJ...671...40R,2010ApJ...714..825G} suggesting that the Cold Spot may be a manifestation of the integrated Sachs-Wolfe effect \citep{1967ApJ...147...73S}, the redshifting of radiation in passing through a void due to the asymmetry of the potential in an accelerating universe. \citet{2015MNRAS.450..288S} have strengthened the case for an extremely large void in depth, or chance superposition of several less extreme voids, in the Cold Spot direction extending over the redshift range $0.05 < z < 0.3$.  \citet{2016MNRAS.462.1882K} argue that the underdense region continues to the relative foreground and they name the feature the Eridanus supervoid.  Several authors \citep{2016MNRAS.455.1246F, 2017arXiv170307894N, 2017arXiv170403814M} have evaluated the possibility that the negative fluctuation in the cosmic microwave background is caused by a vast underdensity in the line-of-sight.  The general consensus disfavors the proposition that line-of-sight voids could be important enough to create the Cold Spot from the integrated Sachs-Wolfe effect alone.

Nevertheless, coincidence or not, a dominant repeller is identified in a direction that overlaps with the Cold Spot in projection.  We evaluate the angular extent and significance of the repeller revealed by the velocity field through the method of constrained realizations mentioned briefly in \S 2.  The density with respect to the mean cosmic density within a sphere of radius $R$ centered at a specified location, $\delta(R) = (\rho(R)-{\bar \rho})/{\bar \rho}$ is derived from the Wiener filter analysis.  The uncertainty in density, $\sigma(R)$, is sampled from 54 constrained realizations.   The basin of the repeller is identified to lie at SGX,SGY,SGZ = [6,500, $-$21,300, $-$6,100]~\kms.  Figure~\ref{fig2} shows the development of the signal-to-noise $-\delta(R)/\sigma(R)$ with increasing spheres of radius $R$.

There is a complexity associated with Fig.~\ref{fig2} that requires explanation.  The location of the repeller at $z \sim 0.077$ lies at the extremity of our data zone (a high density of data to $z \sim 0.05$ with sparse supernova sampling to $z \sim 0.1$).  The solid curve in Fig.~\ref{fig2} illustrates the density signal to r.m.s. fluctuations within the intersection of spheres of increasing radius centered on the repeller basin and a sphere of radius 25,000~\kms\ ($z \sim 0.08$) centered on our position.  The choice of the radius of the sphere centered on us is somewhat arbitrary but roughly captures the region where the Wiener filter density construction carries information that departs from the mean.  In the case of the dashed curve in the figure, the same signal/uncertainty information is illustrated but now within spheres centered on a location in the direction of the Cold Spot.  The specific center is at SGX,SGY,SGZ = [3,600, -16,700, -12,900] \kms, in the direct line-of-sight of the cosmic microwave background Cold Spot with a distance that is the same as that of the repeller basin.  The curve plots $-\delta(R)/\sigma(R)$ within the intersection of spheres centered on the identified location and the 25,000~\kms\ sphere centered on us.  


From the solid curve in Fig.~\ref{fig2} it is seen that the significance of the under density increases with volume, reaching $\sim 2.3\sigma$ by $R \sim 12,000$~\kms.  The peculiar velocity field provides suggestive evidence for an under density on this vast scale.\footnote{The signal in the density field is weak at the extremity of the data where densities from the Wiener filter construction tend to the mean. The potential field is better defined by tidal influences at large distances.}  We note that the direction to the cosmic Cold Spot lies within this domain.  The dashed curve in the figure illustrates the significance of the under density if we shift the assumed center from our preferred location by a distance equivalent to 8,900~\kms\ to a center along the line-of-sight toward the Cold Spot while keeping the same distance.  The significance is reduced but the scale of the implied under density is similar.  

\subsection{Attractors}

The dominant attractor with {\it Cosmicflows-3}, as with {\it Cosmicflows-2}, is coincident with the Shapley Concentration \citep{1989Natur.338..562S,1989Natur.342..251R}.  In detail, there is a shift on the sky of $10^{\circ}$ (to glon=306, glat=+22) and distance of +1,500~\kms\ (to 15,900~\kms).  The new 6dFGSv data \citep{2014MNRAS.445.2677S} implies that the mass overdensity in this region extends into the galactic plane and to the south of the Milky Way, with a secondary maximum near to the south celestial pole.  However, aside from the immediate Shapley region there is a poor correspondence with observed galaxies \citep{2012ApJS..199...26H} or X-ray clusters \citep{2007ApJ...662..224K}.  Again, as with the Cold Spot Repeller, the postulated structure is at the extremity of the data zone.  It has to be suspected that the true nature of the attractor in this region will only be revealed by a study to greater depth.


\section{Discussion}

In their discussion of 6dFRSv results, \citet{2014MNRAS.445.2677S} compared the velocity field implied by their distance and redshift observations with expectations from two independent redshift surveys \citep{1999MNRAS.308....1B,2006MNRAS.373...45E}.  With both comparisons, they noted a large scale departure from the redshift survey expectations, with observed peculiar velocities systematically negative in the Pisces-Cetus sector \citep{1992ApJ...388....9T} and positive in the Centaurus sector, the domain of the Shapley Concentration \citep{1930BHarO.874....9S,1989Natur.338..562S,1989Natur.342..251R}.  
It is a limitation of the redshift survey approach, though, that there is no account of tidal effects from beyond the range of the survey. 

Our approach of inferring mass distribution from velocities accesses information on distant structures from tidal signatures.  Sensitivity to their characteristics diminish with distance.  The repeller in the direction toward the cosmic microwave background Cold Spot is {\it the dominant negative density feature} of the Wiener filter construction with {\it Cosmicflows-3}.  The deviation in direction between the repeller basin and the Cold Spot of $22^{\circ}$ is within the uncertainty of our measurement.  Our distance to the repeller of $\sim 23,000$~\kms\ should only be considered a rough lower limit.

We call the under density identified by the Wiener filter analysis the Cold Spot Repeller because of its coincidence in direction with the Cold Spot fluctuation in the microwave background.  We are providing increased evidence for the existence of a substantial void, or succession of voids, in the Cold Spot direction  \citep{2010ApJ...714..825G, 2015MNRAS.450..288S, 2016MNRAS.462.1882K}.  It remains to be determined if the coincidence in direction has a physical basis.
The tentative nature of the association of our repeller and the cosmic microwave background Cold Spot must be acknowledged.
The observational claim that is made here awaits confirmation with redshift and distance measurements to greater distances.  
However it is likely that proper resolution will require surveys to twice the current depth, with serious completion to $z\sim0.1$, involving an order of magnitude more galaxies or distance estimators with higher accuracy.

\bigskip

\noindent
{\bf Acknowledgements}

Financial support for the Cosmicflows program has been provided by the US National Science Foundation award AST09-08846, an award from the Jet Propulsion Lab for observations with {\it Spitzer Space Telescope}, and NASA award NNX12AE70G for analysis of data from the {\it Wide-field Infrared Survey Explorer}.  Additional support has been provided by the Institut Universitaire de France and the CNES, and the Israel Science Foundation (1013/12).
  
\bibliographystyle{aasjournal}

\end{document}